\documentclass[11pt]{article}

\usepackage{graphicx}

\def \be{\begin{equation}}
\def \ee{\end{equation}}
\def \d{\partial}
\def \aa{\alpha}
\def \bb{\beta}
\def \dd{{\delta}}
\def \gg{\gamma}
\def \ss{\sigma}
\def \ll{\lambda}

\def \kk{\kappa}
\def \DD{\Delta}
\def \SS{\Sigma}
\def \LL{\Lambda}

\def \Mc{{\cal M}}

\def \Tc{{\cal T}}

\def \Vc{{\cal V}}

\def \fr{\frac}

\def \cb{{\bar c}}

\def \gb{{\bar g}}
\def \ggb{{\bar \gg}}

\def \eb{{\bar e}}

\def \aab{{\bar \aa}}
\def \bbb{{\bar \bb}}

\def \psib{{\bar \psi}}

\def \Tt{{\tilde{T}}}
\begin{document}
\title{Lorentz Symmetry Violation in QCD and the Frustration of Asymptotic Freedom}
\author{I.T. Drummond\thanks{email: itd@damtp.cam.ac.uk} \\
            Department of Applied Mathematics and Theoretical Physics\\
            Centre for Mathematical Sciences\\
            Wilberforce Road\\ Cambridge\\ England, CB3 0WA
        }
\maketitle

\abstract{
We study the effect of Lorentz symmetry violation (LSV) on the behaviour at high energy 
of SU(N) gauge theory with quarks in the fundamental representation. The approach 
is similar to that for QED treated in a previous paper. In contrast to QED,
standard Lorentz invariant QCD is asymptotically free. Our aim is to explore the 
structure of the renormalisation group at high energy and hence weak coupling without 
requiring the Lorentz symmetry breaking to be small. The simplest type of LSV leaves
the theory invariant under a subgroup of the Lorentz group that 
preserves a (time-like) 4-vector. We examine this case in detail and find that asymptotic 
freedom is frustrated. 
That is, at sufficiently high energy the running coupling constant attains a minimum value 
before increasing again, while the LSV parameter increases without bound. 
}

\vfill

\pagebreak

\section{\bf Introduction}

Lorentz symmetry violation (LSV) in QED has been studied by a number of authors concerned with
its consistency with causality, unitarity \cite{SCHR,KLINK1,KLINK2}, the structure of asymptotic 
states and renormalisation theory \cite{KOST2,KOST4,LEHN}.
In previous papers \cite{ITD3,ITD4} we studied some of these issues in QED starting with a premetric 
formulation \cite{ITIN,GOMB} based on an action
\be
S=-\fr{1}{8}\int d^4x U^{\mu\nu\ss\tau}F_{\mu\nu}(x)F_{\ss\tau}(x),
\ee
where $F_{\mu\nu}(x)$ is the standard electromagnetic field tensor and the
(constant) background tensor $U^{\mu\nu\ss\tau}$ has the same symmetry properties as
the Riemann tensor in General Relativity, namely
\be
U^{\mu\nu\ss\tau}=-U^{\nu\mu\ss\tau}=U^{\ss\tau\mu\nu},
\ee
and
\be
U^{\mu\nu\ss\tau}+U^{\mu\ss\tau\nu}+U^{\mu\tau\nu\ss}=0.
\ee
This latter condition excludes parity violation. An  outcome of the analysis was that even when
the Lorentz symmetry violation is not constrained to be small the behaviour of the renormalised theory
in the infra-red limit is dominated by the fixed point at zero coupling in a manner consistent
with Lorentz symmetry. That is at a sufficiently large scale in spacetime Lorentz symmetry re-emerges.
This is consistent with related earlier work \cite{NLSN1,NLSN2,NLSN3}. 

In this paper we study a QCD type model with $SU(N)$ gauge symmetry. In
addition to the gauge field we include a quark field that transforms under the fundamental 
representation of $SU(N)$. A closely related model is investigated in reference \cite{APETROV}. 
The significance of such a theory is that it exhibits 
asymptotic freedom, that is, its behaviour at high energy is controlled, at least in the standard case of
Lorentz invariance, by a weak coupling fixed point \cite{GROW,POLZ}. Our aim here is to investigate 
the manner in which asymptotic freedom is modified by the presence of Lorentz symmetry violation.
An investigation with similar aims, in particular comparing QED and QCD is presented in reference \cite{VIEIRA}.
Although we look in detail only at the simplest type of LSV, we set out the general theory
in a manner parallel to reference \cite{ITD3} in order to clarify the logical structure
of the argument. This prepares a framework for analyses of more complex models.     

In the obvious generalisation of the case of QED we take the action for the $SU(N)$ gauge field
to be 
\be
S_{g}=-\fr{1}{8}\int d^4xU^{\mu\nu\ss\tau}F_{a\mu\nu}(x)F_{a\ss\tau}(x),
\label{GINVACT}
\ee
where $F_{a\mu\nu}(x)$ is the standard gauge field tensor transforming according to the
orthogonal representation of $SU(N)$. For a general choice of $U^{\mu\nu\ss\tau}$ this 
action although gauge invariant is not in general Lorentz invariant. Lorentz invariance with respect to
a metric $g^{\mu\nu}$ can be recovered by choosing
\be
U^{\mu\nu\ss\tau}=g^{\mu\ss}g^{\nu\tau}-g^{\nu\ss}g^{\mu\tau}.
\ee
Although there is {\it a priori} no metric in the general case with LSV, there is nevertheless, 
as argued in reference \cite{ITD3}, a {\it preferred} metric $g^{\mu\nu}$ that allows us to 
decompose $U^{\mu\nu\ss\tau}$ in the following way
\be
U^{\mu\nu\ss\tau}=g^{\mu\ss}g^{\nu\tau}-g^{\nu\ss}g^{\mu\tau}-C^{\mu\nu\ss\tau},
\ee
where the tensor $C^{\mu\nu\ss\tau}$ has the same symmetries as the Weyl tensor in General Relativity. That is
\be 
C^{\mu\nu\ss\tau}=-C^{\nu\mu\ss\tau}=C^{\ss\tau\mu\nu},
\ee
and
\be
C^{\mu\nu\ss\tau}+C^{\mu\ss\tau\nu}+C^{\mu\tau\nu\ss}=0.
\ee
In addition it satisfies the trace condition
\be
g_{\mu\ss}C^{\mu\nu\ss\tau}=0.
\ee
We refer to $C^{\mu\nu\ss\tau}$ as a Weyl-like tensor (WLT). It follows that the WLT determines the 
nature of the LSV. As in the case of QED the possible types of LSV can be determined by applying
the Petrov classification to the WLT \cite{PTRV}. A useful approach to the Petrov scheme is contained
in references \cite{JMS,PODON}. Its application in QED with LSV is presented in reference \cite{ITD3}. There are six
cases, conventionally labeled O,N,D,I,II,III. Each case has a canonical form for the WLT \cite{PENRIN}.
 
Class O corresponds to the case $C^{\mu\nu\ss\tau}=0$ which for pure gauge theory implies no
LSV. However as in the case of QED \cite{ITD3}, the quark field can engender LSV in the model
through its contribution to vacuum polarisation provided the associated metric for quark propagation
shares with the gluon metric an invariance under a subgroup of the Lorentz group that is the little group
of the given 4-vector \cite{COLGL1}. The 4-vector can be time-like, space-like or light-like 
(with respect to both metrics). The time-like case implies that there is a reference frame in 
which the theory is invariant under rotations of the spatial axes. This is the case we study in detail. 
However it is convenient to set out the scheme for quantising and renormalising the theory in a general 
form. Canonical forms for the WLT in other Petrov classes and the implications for the vector meson 
dispersion relations are the same as those for photons in QED as described in detail in reference \cite{ITD3}.

\section{\label{GFIX} Gauge Fixing and Ghost Fields}

In terms of the vector gauge fields $A_{a\mu}(x)$ the tensor fields are given by
\be
F_{a\mu\nu}(x)=\d_\mu A_{a\nu}(x)-\d_\nu A_{a\mu}(x)+gf_{abc}A_{b\mu}(x)A_{c\nu}(x),
\ee
where $g$ is the gauge field coupling constant and $f_{abc}$ are the structure constants of SU(N).
In order to deal with the gauge invariance of the action for the vector fields in eq(\ref{GINVACT})
we follow the approach in reference \cite{ITD3} and impose the gauge condition
\be
\LL^{\mu\nu}\d_\mu A_{a\nu}(x)=0.
\ee
Here $\LL^{\mu\nu}$ is a metric-like tensor which we will find it convenient to distinguish from $g^{\mu\nu}$
because the two tensors behave differently under the renormalisation procedure.
We are therefore led to add a gauge fixing term to the action of the form

\be 
S_{gf}=\fr{1}{2}\int d^4x(\LL^{\mu\nu}\d_\mu A_{a\nu}(x))^2.
\label{GFACT}
\ee

In addition and in contrast to the case of QED \cite{ITD3}, we must introduce anticommuting ghost fields 
$c_a(x)$ and $\cb_a(x)$ in order to construct in the standard way the Fadeev-Popov determinant in the 
path integral formalism for the computation of Greens functions in the gauge theory. We therefore complete the 
action for the gauge theory by adding a term 
\be
S_{gh}=-\int d^4x \cb_a(x)\LL^{\mu\nu}\d_\mu D_{ab\nu}c_b(x).
\label{GSTACT}
\ee
Here $D_{ab\nu}=\dd_{ab}\d_\nu+gf_{abc}A_{c\nu}(x)$ is the gauge covariant derivative for the ghost fields.
The complete action for the theory is $S$ where

\be
S=S_{g}+S_{gf}+S_{gh}.
\label{TOTACT}
\ee

\section{\label{FEYN} Feynman Rules}

The Feynman rules for the theory can be read off from the action $S$ in the standard way.
They are in certain respects analogous to the corresponding rules for BIMQED \cite{ITD3}.

\subsection{\label{PROP} Feynman Propagator}

\begin{figure}[t]
  \centering
  \includegraphics[width=0.6\linewidth]{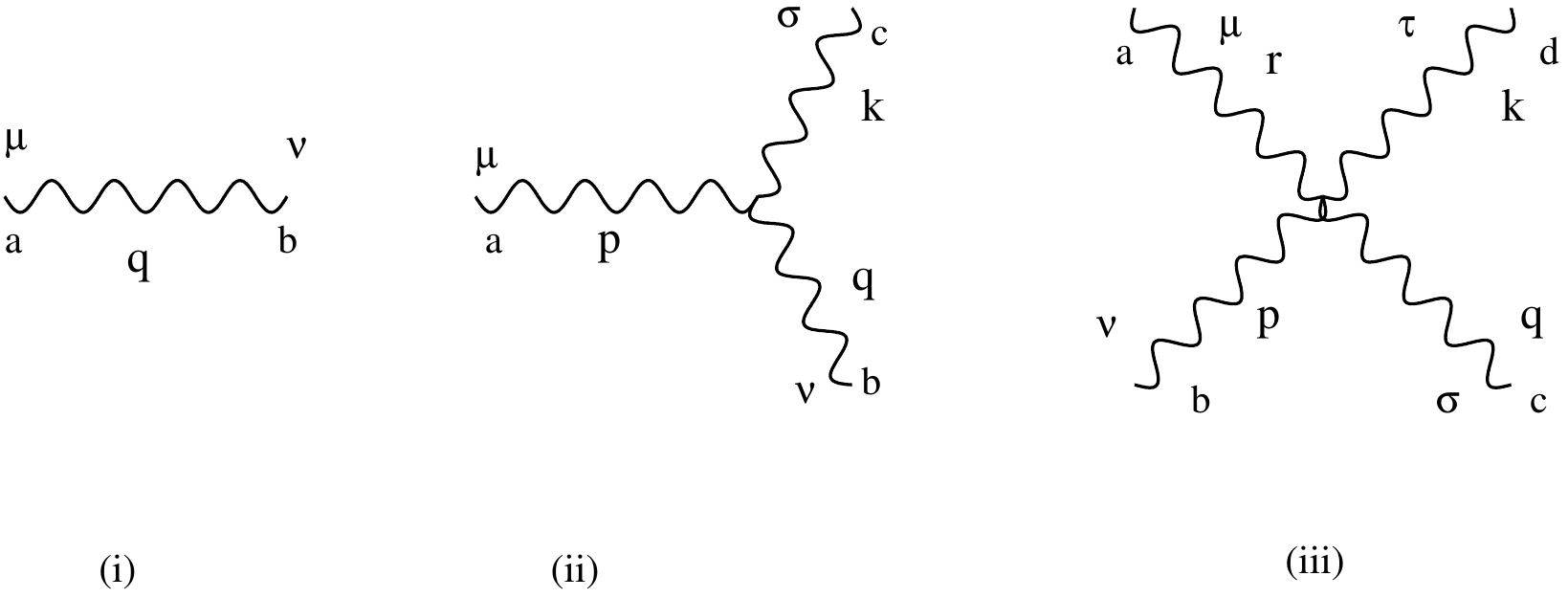}
  \caption{Gluons (i) propagator, (ii) 3-vertex, (iii) 4-vertex. 4-Momenta inward.}
  \label{FIG1}
\end{figure}

The Feynman propagator, illustrated in Fig \ref{FIG1}(i), is
\be
\DD_{Fab\mu\nu}(q)=-i\dd_{ab}M_{\mu\nu}(q),
\ee
where $M_{\mu\nu}(q)$ is the matrix inverse to $M^{\mu\nu}(q)$ and
\be
M^{\mu\nu}(q)=(U^{\mu\aa\nu\bb}+\LL^{\mu\aa}\LL^{\nu\bb})q_{\aa}q_{\bb}.
\ee
More explicitly
\be
M^{\mu\nu}(q)=q^2g^{\mu\nu}-q^\mu q^\nu+Q^\mu Q^\nu -C^{\mu\aa\nu\bb}q_\aa q_\bb.
\ee
Here $Q^\mu=\LL^{\mu\aa}q_{\aa}$. It is easy to verify that when $C^{\mu\aa\nu\bb}$
vanishes and $\LL^{\mu\nu}=g^{\mu\nu}$ this reduces to the standard Lorentz
invariant form. Following the analysis for the photon propagator in reference \cite{ITD3},
we first introduce $\Mc^{\mu\nu}(q)$ which is the form taken by $M^{\mu\nu}(q)$ when indeed $\LL^{\mu\nu}$
is replaced by $g^{\mu\nu}$ that is 

\be
\Mc^{\mu\nu}(q)=q^2g^{\mu\nu}-C^{\mu\aa\nu\bb}q_{\aa}q_{\bb}.
\ee
The inverse matrix is $\Mc_{\mu\nu}(q)$ (see reference \cite{ITD3}) and it can be used to construct $M_{\mu\nu}(q)$
in the form

\be
M_{\mu\nu}(q)=\left(\dd^\ss_\mu-\fr{q_\mu Q^\ss}{q.Q}\right)\Mc_{\ss\tau}(q)
                                          \left(\dd^\tau_\nu-\fr{Q^\tau q_\nu}{q.Q}\right)
                                                                          +\fr{q_\mu q_\nu}{(q.Q)^2}
\ee
A careful analysis \cite{ITD3} shows that the Feynman propagator has the same vector meson poles
as $\Mc_{\mu\nu}(q)$ together with the ghost poles at $q.Q=\LL^{\mu\nu}q_\mu q_\nu=0$. It is straightforward
to check that were we to set $\LL^{\mu\nu}=g^{\mu\nu}$ then we find
\be
M_{\mu\nu}(q)=\Mc_{\mu\nu}(q)
\ee

\subsection{\label{VERTS} Gluon Vertices}

The three-gluon vertex, Fig \ref{FIG1}(ii), is
\be
V^{\mu\nu\ss}_{abc}=-gf_{abc}(p_\rho U^{\rho\mu\nu\ss}+q_\rho U^{\rho\nu\ss\mu}+k_\rho U^{\rho\ss\mu\nu}).
\ee
For the four-gluon vertex Fig \ref{FIG1}(iii) we have 
\be
V^{\mu\nu\ss\tau}=-ig^2(U^{\mu\nu\ss\tau}f_{hab}f_{hcd}+U^{\mu\ss\tau\nu}f_{hac}f_{hbd}
                             +U^{\mu\tau\nu\ss}f_{had}f_{hbc}).
\ee
Of course momentum conservation is enforced at each vertex. Again it is easy to verify that 
in the absence of LSV these vertices reduce to standard form ({\it see for example}
\cite{PESK}).

\subsection{\label{GHOST} Ghost Propagator and Vertex}

\begin{figure}[t]
 \centering
 \includegraphics[width=0.4\linewidth]{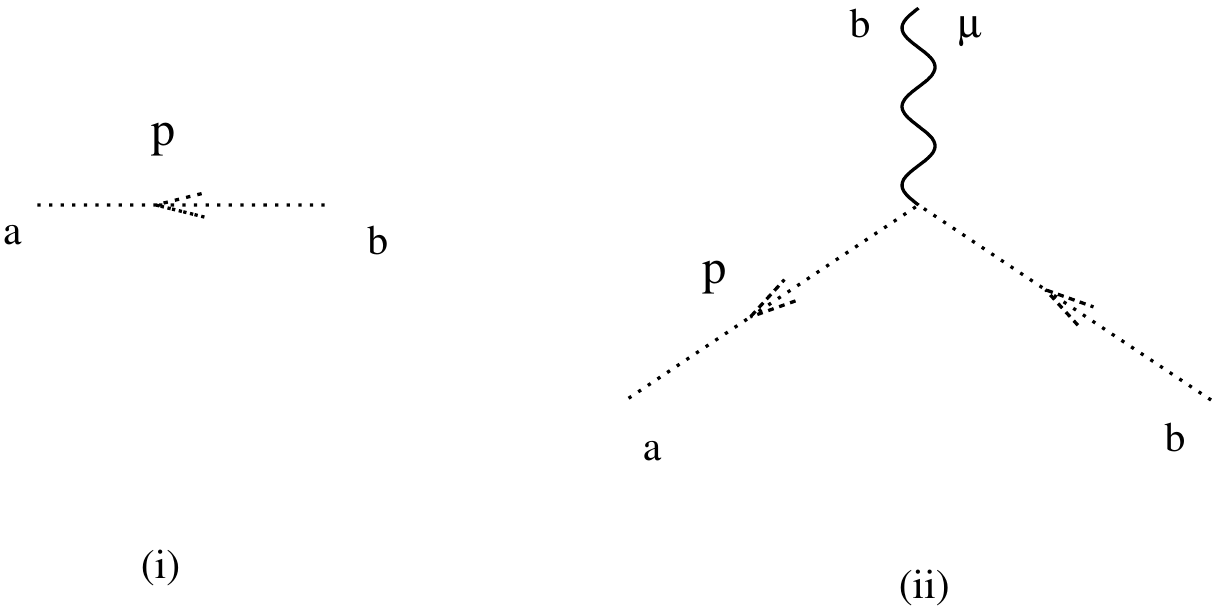}
 \caption{The ghost propagator is indicated in (i) and the vertex for coupling to gluons in (ii).}
 \label{FIG2}
\end{figure}

The Feynman propagator for the ghost fields, Fig \ref{FIG2}(i), is
\be
\DD^{(gh)}_{ab}(p)=i\fr{\dd_{ab}}{P.p},
\ee
where $P^\mu=\LL^{\mu\nu}p_{\nu}$ and $P.p=P^\mu p_\mu=\LL^{\mu\nu}p_\mu p_{\nu}$. 
Momentum follows the ghost direction.
From this it is obvious that the mass-shell condition for the ghosts is $P.p=\LL^{\mu\nu}p_\mu p_{\nu}=0$.

The vertex coupling the ghosts to the vector field is indicated in Fig \ref{FIG2}(ii) 
and has the form
\be
V^{(gh)\mu}_{abc}=-gf_{abc}\LL^{\mu\nu}p_\nu.
\ee

\section{\label{QUARK} Quark Field}

\begin{figure}[t]
 \centering
 \includegraphics[width=0.4\linewidth]{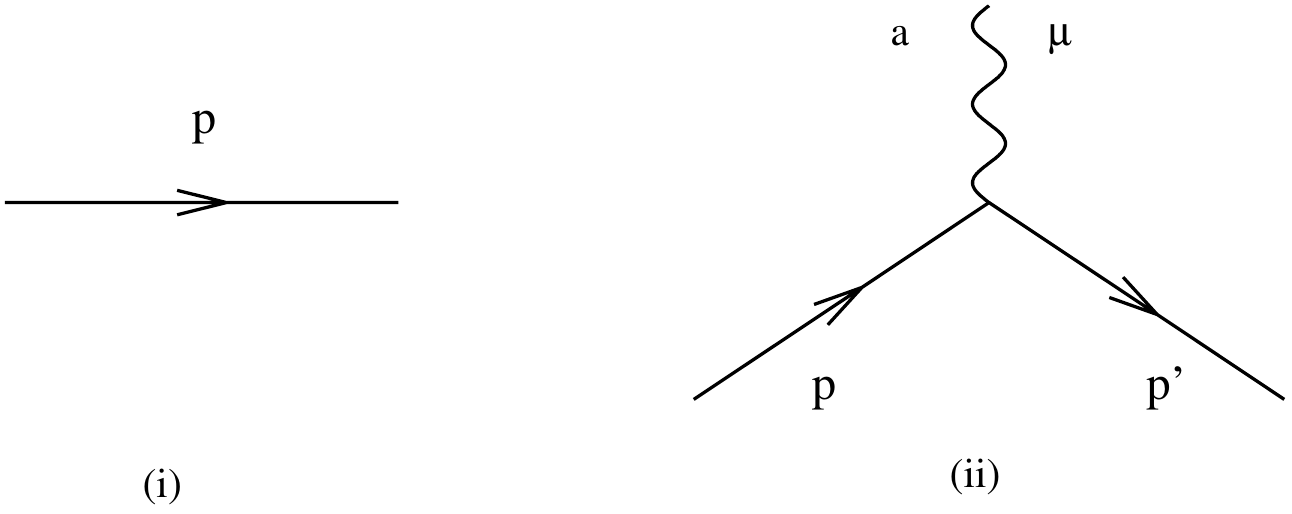}
 \caption{The quark propagator is indicated in (i) and the quark gluon coupling in (ii).}
 \label{FIG3}
\end{figure}

The model can be extended by including one or more spinor fields each transforming under $SU(N)$.
For simplicity we will consider the case of one such field. Modifications can be added later.
The action for this field is $S_{qu}$ where
\be
S_{qu}=\int d^4x\psib(x)(i\ggb^\mu D_\mu[A]-m)\psi(x).
\ee
Here
\be
\ggb^\mu=\gg^a\eb^\mu_{~~a},
\ee
where $\eb^\mu_{~~a}$ is the vierbein associated with the spinor field and $\{\gg^a\}$ are the standard
Dirac $\gg$-matrices. The metric associated with the spinor field is $\gb^{\mu\nu}=\eta^{ab}\eb^\mu_{~~a}\eb^\nu_{~~b}$.
This is a second source of LSV in the model.
Of course $D_\mu[A]$ is the gauge covariant derivative appropriate to the spinor field. The Feynman propagator
for the spinor field is indicated in Fig \ref{FIG3}(i)
\be
S_F(p)=\fr{i}{\ggb^\mu p_\mu-m},
\ee
and the coupling to the gauge field is indicated in Fig \ref{FIG3}(ii)
\be
\Vc^\mu_a=ig\ggb^\mu t_a.
\ee
Here $t_a$ is the $SU(N)$ generator appropriate to the representation of the spinor field
(see reference \cite{PESK}). 

We use dimensional regularisation \cite{tHV} in order to deal with the ultraviolet divergences of the theory.
The Feynman diagrams for the perturbation series for the Greens functions of the theory are evaluated in 
$n$-dimensions. All the parameters of the theory, coupling constant, mass, metrics and WLT
are subject to corrections involving UV divergences. We denote the bare quantities with an 
extra suffix $0$. Each bare parameter is expanded in terms of a renormalised coupling constant $g$. For example
\be
g\rightarrow g_0=\mu^{(4-n)/2}g(1+\sum_{k=1}^{\infty}g^{(k)}g^{2k}).
\label{RENORM1}
\ee
Here $\mu$ is the scale associated with the renormalised coupling $g$ \cite{tHV}. The coefficients $g^{(k)}$
depend on the dimension $n$ and exhibit poles of various orders at $n=4$. For example $g^{(1)}$ 
has a simple pole at  $n=4$. The other parameters are similarly replaced by bare versions that are 
expanded in powers of the renormalised coupling $g$.
\begin{eqnarray}
 m_0&=&m(1+\sum_{k=1}^\infty b^{(k)}g^{2k})\nonumber\\
 g_0^{\mu\nu}&=&g^{\mu\nu}+\sum_{k=1}^\infty g^{(k)\mu\nu}g^{2k}\nonumber\\
 \eb^\mu_{0~a}&=&\eb^\mu_{~~a}+\sum_{k=1}^\infty \eb^{(k)\mu}_{~~~~~a}g^{2k}\nonumber\\
 C_0^{\mu\nu\ss\tau}&=&C^{\mu\nu\ss\tau}+\sum_{k=1}^\infty C^{(k)\mu\nu\ss\tau}g^{2k}\nonumber\\
 \LL_0^{\mu\nu}&=&\LL^{\mu\nu}+\sum_{k=1}^\infty \LL^{(k)\mu\nu}g^{2k}
\label{RENORM2}
\end{eqnarray}

Again the coefficients in the various expansions exhibit poles at $n=4$. Note that we are free to
assume that $\det g_0^{\mu\nu}=\det g^{\mu\nu}=-1$. This implies that
\be
g_{\mu\nu}g^{(1)\mu\nu}=0.
\label{RENORM3}
\ee

\section{\label{LOOP1} Perturbative Calulations at One-loop}

\subsection{\label{GPROP} Renormalisation of Gauge Field Propagator}

The vacuum polarisation tensor $\SS^{\mu\nu}_{ab}(q)$ determines the renormalisation properties of
the (inverse) gluon propagator $\DD^{\mu\nu}_{ab}(q)$. The (one-loop) diagrams that contribute to 
$i\SS^{\mu\nu}_{ab}(q)$  are shown in Fig \ref{FIG4} (see reference \cite{PESK}). We have, to $O(g^2)$
\be
\DD^{\mu\nu}_{0ab}(q)=\DD^{\mu\nu}_{F0ab}(q)+i\SS^{\mu\nu}_{ab}(q).
\label{RENORM4}
\ee
Here the inverse Feynman propagator is expressed in terms of the bare parameters,
\be
\DD^{\mu\nu}_{F0ab}(q)=-i\dd_{ab}\left\{(g_0^{\mu\nu}g_0^{\aa\bb}-g_0^{\mu\bb}g_0^{\nu\aa}+\LL_0^{\mu\bb}\LL_0^{\nu\aa}
                               -C_0^{\mu\aa\nu\bb})q_\aa q_\bb\right\}.
\label{RENORM5}
\ee
In principle the contributions to the vacuum polarisation are also computed from the appropriate
Feynman diagrams using the bare parameters. However our calculation will be of $O(g^2)$
so we need only use the lowest order expansions in eqs(\ref{RENORM2}). This amounts to using the 
renormalised parameters in the vertices and propagators when computing $i\SS^{\mu\nu}_{ab}(q)$. In
addition the UV divergences of $i\SS^{\mu\nu}_{ab}(q)$ occur only in the lowest terms in 
its Taylor expansion in $q_\aa$. On general grounds then we can exhibit the UV divergences to $O(g^2)$
by writing
\be
i\SS^{\mu\nu}_{ab}(q)=i\dd_{ab}\fr{g^2}{n-4}W^{\mu\aa\nu\bb}q_\aa q_\bb+O(q^4),
\label{RENORM6}
\ee
where the $O(q^4)$ terms are UV-finite. The tensor $W^{\mu\aa\nu\bb}$ has the same symmetry properties as 
the Riemann tensor. Hence, in a standard fashion, it can be expressed in the form
\be
W^{\mu\aa\nu\bb}=\left\{\fr{1}{12}W(g^{\mu\nu}g^{\aa\bb}-g^{\mu\bb}g^{\nu\aa})
               +\fr{1}{2}(V^{\mu\nu}g^{\aa\bb}+g^{\mu\nu}V^{\aa\bb}-V^{\mu\bb}g^{\nu\aa}-g^{\mu\bb}g^{\nu\aa})
                     -V^{\mu\aa\nu\bb}\right\},
\label{RENORM7}
\ee 
where 
\begin{eqnarray}
W&=&W^{\aa\bb}g_{\aa\bb},\nonumber\\
W^{\aa\bb}&=&W^{\mu\aa\nu\bb}g_{\mu\nu},\nonumber\\
V^{\aa\bb}&=&W^{\aa\bb}-\fr{1}{4}Wg^{\aa\bb}.
\label{WEYL1}
\end{eqnarray}
We have also
\begin{eqnarray}
V^{\aa\bb}g_{\aa\bb}&=&0,\nonumber\\
V^{\mu\aa\nu\bb}g_{\aa\bb}&=&0.
\label{WEYL2}
\end{eqnarray}
It follws that $V^{\mu\aa\nu\bb}$, having the appropriate symmetries and trace properties,  
is a WLT. In equations (\ref{WEYL1}) and (\ref{WEYL2}) we have used 
the 4-D decomposition which is adequate when computing the residues of the pole at $n=4$. The type 
of Lorentz symmetry breaking exhibited by the model can be specified by means of the Petrov 
classification of Weyl tensors. We will return to this point later.
 
From equations (\ref{RENORM1}), (\ref{RENORM2}) and (\ref{RENORM5}) we see that
\be
\DD_{F0ab}^{\mu\nu}(q)=\DD_{Fab}^{\mu\nu}(q)+\dd\DD_{Fab}^{\mu\nu}(q),
\label{RENORM8}
\ee
where $\DD_{Fab}^{\mu\nu}(q)$ is obtained from $\DD_{F0ab}^{\mu\nu}(q)$ by replacing the bare parameters
with their renormalised versions and
\begin{eqnarray}
\dd\DD_{Fab}^{\mu\nu}(q)&=&-\dd_{ab}ig^2\{g^{\aa\bb}g^{(1)\mu\nu}+g^{(1)\aa\bb}g^{\mu\nu}
                          -g^{\mu\aa}g^{(1)\nu\bb}-g^{(1)\mu\aa}g^{\nu\bb}\nonumber\\
&&            ~~~~~~~~~~~~        +\LL^{\aa\bb}\LL^{(1)\mu\nu}+\LL^{(1)\aa\bb}\LL^{\mu\nu}
                           -\LL^{\mu\aa}\LL^{(1)\nu\bb}-\LL^{(1)\mu\aa}\LL^{\nu\bb}\nonumber\\
&&            ~~~~~~~~~~~~       -C^{(1)\mu\aa\nu\bb}\}q_\aa q_\bb
\label{RENORM9}
\end{eqnarray}
The renormalisation parameters are fixed by requiring that the renormalisation of $\DD_{0ab}^{\mu\nu}(q)$
reduces to an overall multiplicative factor, that is
\be
\DD_{0ab}^{\mu\nu}(q)=\left(1-\fr{1}{12}\fr{g^2}{n-4}W\right)\left(\DD_{Fab}^{\mu\nu}(q)+O(q^4)\right).
\label{RENORM10}
\ee
This is achieved by requiring that

\begin{eqnarray}
g^{(1)\mu\nu}&=&\fr{1}{2}\fr{1}{n-4}V^{\mu\nu},\nonumber\\
\LL^{(1)\mu\nu}&=&\fr{1}{24}\fr{1}{n-4}W\LL^{\mu\nu},\nonumber\\
C^{(1)\mu\aa\nu\bb}&=&\fr{1}{n-4}\left(V^{\mu\aa\nu\bb}-C^{\mu\aa\nu\bb}\right).
\label{RENORM11}
\end{eqnarray}
Note that the results in equations(\ref{RENORM11}) do imply that $g^{(1)\mu\nu}g_{\mu\nu}=0$.
Furthermore the renormalisation of $\LL_0^{\mu\nu}$ is multiplicative and involves the field 
renormalisation factor $Z^{1/2}$ where
\be 
Z^{1/2}=1+\fr{1}{24}\fr{g^2}{n-4}W.
\label{RENORM12}
\ee
This is not necessarily true of the metric renormalisation. For this reason it is conceptually
convenient to distinguish the bare metric from the bare ghost metric. However at the level of
one loop in perturbation theory it is possible and convenient to allow the two metrics to
coincide. Higher order calculations may require the maintenance of the  distinction, imposed
order by order, between the two metrics.

\subsection{\label{QPROP} Renormalisation of Quark Propagator}

The renormalisation of the quark propagator $iS_0(p)$ proceeds along similar lines. We have

\be
S_0^{-1}(p)=S_{F0}^{-1}(p)+\SS(p),
\label{RENORM13}
\ee
where

\be
S_{F0}^{-1}(p)=\ggb_0^\mu p_\mu+m_0,
\label{RENORM14}
\ee

\be
\ggb_0^\mu=\gg^a e_{0~a}^\mu
\label{RENORM15}
\ee
and $\gg^a$ are standard Dirac $\gg$-matrices. The quark propagator is implicitly 
a unit operator in $SU(N)$ space. The self-energy amplitude $i\SS(p)$
can be calculated (at one loop) from the diagram in Fig \ref{FIG5} using the Feynman rules with
renormalised parameters. The UV divergences can be exposed in the Taylor expansion

\be
\SS(p)=\SS(0)+\SS^\mu(0)p_\mu+O(p^2),
\label{RENORM16}
\ee
where the contribution $O(p^2)$ is finite at $n=4$, and we have
\be
\SS(0)=m\fr{g^2\ss}{n-4},
\label{RENORM17a}
\ee
and
\be
\SS^\mu(0)=\fr{g^2}{n-4}H^\mu_{~~\nu}\ggb^\nu.
\label{RENORM17}
\ee
We write
\be
H^\mu_{~~\nu}=\fr{1}{n}H\dd^\mu_\nu+h^\mu_{~\nu},
\label{RENORM18}
\ee
where $h^\mu_{~~\mu}=0$. The term $H$ determines the quark propagator renormalisation $Z_q$,
while $h^\mu_{~~\nu}$ fixes the counter-term in $\eb^\mu_{0~a}$. More explicitly
we have, taking account only of the poles at $n=4$,
\be
Z_q=1-\fr{1}{4}\fr{g^2}{n-4}H.
\label{RENORM19}
\ee
\be
g^2b^{(1)}=\fr{g^2}{n-4}\ss+\fr{1}{4}\fr{g^2}{n-4}H,
\label{RENORM20}
\ee
and
\be
g^2\eb^{(1)\mu}_{~~~~b}\eb^b_{~~\nu}+\fr{g^2}{n-4}h^\mu_{~~\nu}=0.
\label{RENORM21}
\ee
That is
\be
g^2\eb^{(1)\mu}_{~~~~a}+\fr{g^2}{n-4}h^\mu_{~~\nu}\eb^\nu_{~~a}=0,
\label{RENORM22}
\ee
implying that, to $O(g^2)$
\be
\gb^{\mu\nu}_0=\gb^{\mu\nu}-\fr{g^2}{n-4}(h^\mu_{~~\rho}\gb^{\rho\nu}+h^\nu_{~~\rho}\gb^{\rho\mu}).
\label{RENORM23}
\ee

\subsection{\label{COUP} Renormalisation of Coupling Constant}

The coupling constant renormalisation is most easily followed by considering the 
quark-gluon vertex. The relevant diagrams are shown in Fig \ref{FIG6} and
yield contributions to the truncated three-point function that render it finite
after appropriate field renormalisations. We can anticipate the nature 
of the divergences by examining the bare vertex $\Vc^\mu_{0a}=ig_0\ggb_0^\mu t_a$.
From eqs (\ref{RENORM1}) and (\ref{RENORM2}) we have
\be
\Vc^\mu_{0a}=ig_0\ggb_0^\mu t_a
               =i\mu^{(4-n)/2}g(1+g^2g^{(1)})(\eb^{\mu}_{~~b}+g^2\eb^{(1)\mu}_{~~~~b}))\gg^b t_a.
\label{RENORM24}
\ee
That is
\be
\Vc^\mu_{0a}=i\mu^{(4-n)/2}g\left(\left(1+g^2g^{(1)}\right)\dd^\mu_\nu-\fr{g^2}{n-4}h^\mu_{~~\nu}\right)\ggb^\nu t_a.
\label{RENORM25}
\ee
The point we make here is that the renormalisation of the quark vierbein enters 
into the vertex calculation and the contribution $\Vc^\mu_a(p,p')$ to the three point function from 
the diagrams in Fig \ref{FIG6} must be consistent with this. We can expect then that $\Vc^\mu_a(p,p')$ will (at zero
external momenta) have the form
\be
\Vc^\mu_a(0,0)=i\mu^{(4-n)/2}g\left(\fr{g^2}{n-4}K\dd^\mu_\nu t_a+i\fr{g^2}{n-4}h^\mu_{~~\nu}\right)\ggb^\nu t_a.
\label{RENORM26}
\ee
This will be verified in particular calculations.
The {\it truncated} three point function $\Vc^{(3)\mu}_a(p,p')$ will, at zero external momentum, satisfy
\be
\Vc^{(3)\mu}_a(0,0)=i\mu^{(4-n)/2}g\left(1+g^2g^{(1)}+\fr{g^2}{n-4}K\right)\ggb^\mu t_a.
\label{RENORM27}
\ee
Finally $g^{(1)}$ is determined by requiring that the right side of this equation is rendered 
finite by extracting the field renormalisation factors $Z^{-1}Z_q^{-1/2}$, implying 
\be
g^2g^{(1)}=\fr{g^2}{n-4}\left(\fr{1}{4}H-\fr{1}{24}W-K\right).
\label{RENORM28}
\ee
We will look at this in more detail when evaluating the vertex
in the special case of LSV we consider below.

\section{\label{RENGRP} Renormalisation Group}

From the results in section \ref{LOOP1} we can obtain the renormalisation group 
equations for the renormalised parameters to lowest non-trivial order in the 
coupling constant $g$. These are derived from the requirement that the bare parameters
are indepependent of the renormalisation scale $\mu$. For the coupling constant we have
\be
\mu\fr{\d}{\d\mu}g_0=0.
\label{RENGRP1}
\ee
From eq(\ref{RENORM26}) we then obtain the renormalisation group $\bb$-function
\be
\bb(\mu)=\mu\fr{\d}{\d\mu}g=-(2-n/2)g-g^3\left(\fr{1}{4}H-\fr{1}{24}W-K\right).
\label{RENGRP2}
\ee
Note that in deriving eq(\ref{RENGRP2}) from eq(\ref{RENGRP1}) we have ignored 
derivatives of $H$, $W$ and $K$ since they are of $O(g^2)$ and may, and indeed must, 
be ignored at one loop. Of course the first term on the right of eq(\ref{RENGRP2})
vanishes in four dimensions. 

Following the same principles we obtain for the gluon metric
\be
\mu\fr{\d}{\d\mu}g^{\aa\bb}=-\fr{1}{2}g^2V^{\aa\bb},
\label{RENGRP3}
\ee
and
\be
\mu\fr{\d}{\d\mu}\gb^{\aa\bb}=g^2(h^\aa_{~~\rho}\gb^{\rho\bb}+h^\bb_{~~\rho}\gb^{\rho\aa}).
\label{RENGRP4}
\ee
We obtain also for the renormalised mass
\be
\mu\fr{\d m}{\d\mu}=-mg^2(\ss+\fr{1}{4}H).
\label{RENGRP5}
\ee

The renormalisation scheme set out here can be applied to LSV associated with a WLT of any of the
Petrov classes. However even at one loop the calculations are rather complex. Partly then for reasons
of simplicity we restrict attention in this paper to models in Petrov class O. The results 
nevertheless remain interesting.

\section{\label{PETROV0} Renormalisation for Petrov Class O} 

As explained above, in the case of Petrov class O, the LSV is due entirely to the 
difference between the light-cone associated with the gluons and that associated with the 
quark field. That is, we are assuming that the tensor $C^{\mu\nu\ss\tau}$ vanishes. This is possible if
there is a reference frame in which rotational invariance holds
simultaneously for both metrics. Given the symmetry properties of $C^{\mu\nu\ss\tau}$ it
is consistent with this rotational invariance only if it is null. 
Similar remarks apply to situations where the preserved subgroup of the Lorentz group, 
in a suitable frame, leaves invariant either a purely space-like or light-like vector. 
We will concentrate on the rotationally invariant case. We have for the gluon metric

\be
g^{\mu\nu}=\left(\begin{array}{cccc}
                      \aa&0&0&0\\
                      0&-\bb&0&0\\
                      0&0&-\bb&0\\
                      0&0&0&-\bb \end{array}\right),
\ee
and for the quark metric
\be\gb^{\mu\nu}=\left(\begin{array}{cccc}
                      \aab&0&0&0\\
                      0&-\bbb&0&0\\
                      0&0&-\bbb&0\\
                      0&0&0&-\bbb \end{array}\right),
\ee
We assume that 
\be
\aa\bb^3=\aab\bbb^3=1, 
\ee
so that $\det g^{\mu\nu}=\det \gb^{\mu\nu}=-1$. We have similar forms for
the bare metrics $g_0^{\mu\nu}$ and $\gb_0^{\mu\nu}$ in terms of the bare parameters
($\aa_0,\bb_0$) and ($\aab_0,\bbb_0$) which have appropriate expansions in powers of the 
renormalised coupling. That is

\be
\aa_0=\aa+\fr{g^2}{n-4}\aa^{(1)}+\ldots,
\ee
together with similar expansions for the other bare parameters. 
It is convenient to relate the two metrics by setting $\aab=a\aa$ and $\bbb=b\bb$ with the
consequence that $ab^3=1$. Similar remarks hold in an obvious way for bare parameters $a_0$ and $b_0$.
The significance then of $b$ is that in a coordinate frame in which the gluon metric
is diagonal with entries$(1,-1,-1,-1)$ the quarks have a lightcone associated with a velocity $c_q=b^2$.
At appropriately high energies we would expect that when $b>1$ the quarks travel faster than the gluons
and slower when $b<1$. 

\subsection{\label{VACPOL0} Vacuum Polarisation for Petrov Class O}

\begin{figure}[t]
 \centering
 \includegraphics[width=0.6\linewidth]{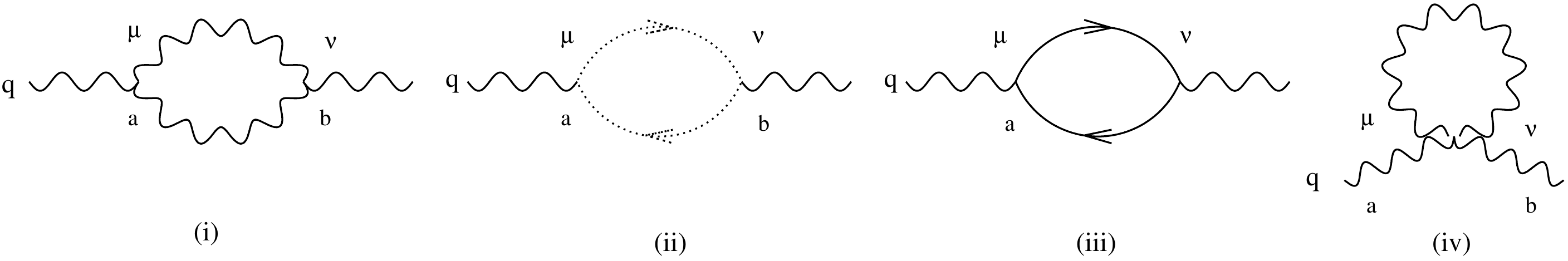}
 \caption{Contributions to vacuum polarisation from (i) gluons, (ii) ghosts, (iii) quarks and (iv) gluon loop.}
 \label{FIG4}
\end{figure}

In order to carry out the renormalisation process we assume the renormalised gauge metric
satisfies (to one loop order) $\LL^{\mu\nu}=g^{\mu\nu}$. The vanishing of $C^{\mu\nu\ss\tau}$ then yields
for the renormalised gluon propagator
\be
M_{\mu\nu}(q)=\fr{g_{\mu\nu}}{q^2}.
\ee
The vertex factors, as mentioned above, acquire standard form for the gauge theory without (at this
stage) any LSV. We can therefore use the discussion of non-abelian gauge theories in  reference \cite{PESK} 
to evaluate 
the contributions from the diagrams in Fig \ref{FIG4}(i), Fig \ref{FIG4}(ii) and Fig \ref{FIG4}(iv)
to obtain the gauge field contribution to the UV divergence at one loop as
\be
i\SS^{(g)\mu\nu}_{ab}(q)=-\fr{10}{3}C_2(G)\fr{ig^2}{(4\pi)^2}\dd_{ab}
                     (g^{\mu\nu}g^{\aa\bb}-g^{\mu\bb}g^{\aa\nu})q_\aa q_\bb\fr{1}{n-4},
\ee
where $C_2(G)$ is the value of the quadratic Casimir operator in the adjoint representation of $SU(N)$.
As explained in reference \cite{PESK} the diagram in Fig \ref{FIG4}(iv) does not contribute to this pole residue.
In a similar way, we can use the results in reference \cite{PESK} to evaluate the quark contribution to the 
vacuum polarisation from Fig \ref{FIG4}(iii) provided we use $\gb^{\mu\nu}$ as the appropriate metric.
The result is
\be
i\SS^{(q)\mu\nu}_{ab}(q)=\fr{4}{3}\fr{ig^2}{(4\pi)^2}\dd_{ab}(\gb^{\mu\nu}\gb^{\aa\bb}-\gb^{\mu\bb}\gb^{\aa\nu})
                                      \fr{1}{n-4}.
\ee
From eq(\ref{WEYL1}) we see that contribution to the tensor $W^{\mu\aa\nu\bb}$ from the gluon field term is
\be
W^{(g)\mu\aa\nu\bb}=-\fr{1}{(4\pi)^2}\fr{10}{3}C_2(G)(g^{\mu\nu}g^{\aa\bb}-g^{\mu\aa}g^{\bb\nu}),
\ee
and
\be
W^{(g)}=-\fr{40}{(4\pi)^2}C_2(G).
\ee
It is immediately obvious $V^{(g)\mu\nu}$ vanishes as does $V^{(g)\mu\aa\nu\bb}$.
From the quark field term we have
\be
W^{(q)\mu\aa\nu\bb}=\fr{1}{(4\pi)^2}\fr{4}{3}(\gb^{\mu\nu}\gb^{\aa\bb}-\gb^{\mu\bb}\gb^{\aa\nu}).
\ee
We have then
\be
W^{(q)}=\fr{8}{(4\pi)^2}\fr{\bbb}{\bb}\left(\fr{\aab}{\aa}+\fr{\bbb}{\bb}\right)
                                             =\fr{8}{(4\pi)^2}b(a+b),
\label{PERT1}
\ee
and
\be
V^{(q)\aa\bb}=\fr{2}{(4\pi)^2}b(a-b)\left(\begin{array}{cccc}
                                                          \aa&0&0&0\\
                                                           0&\bb/3&0&0\\
                                                           0&0&\bb/3&0\\
                                                           0&0&0&\bb/3 \end{array}\right).
\label{PERT2}
\ee
It follows that 
\be
W=W^{(g)}+W^{(q)}=\fr{8}{(4\pi)^2}(b(a+b)-5C_2(G)).
\label{PERT3}
\ee
Of course
\be
V^{\mu\nu}=V^{(q)\mu\nu}.
\label{PERT4}
\ee
Finally then from eqs(\ref{RENORM11}) and (\ref{PERT4}) we have
\be
\aa^{(1)}=\fr{1}{(4\pi)^2}\fr{1}{n-4}b(a-b)\aa=\fr{1}{(4\pi)^2}\fr{1}{n-4}\left(\fr{1}{b^2}-b^2\right)\aa,
\ee
that is
\be
\aa_0=\aa\left(1+\fr{g^2}{(4\pi)^2}\fr{1}{n-4}\left(\fr{1}{b^2}-b^2\right)\right),
\ee
and
\be
\bb_0=\bb\left(1-\fr{1}{3}\fr{g^2}{(4\pi)^2}\fr{1}{n-4}\left(\fr{1}{b^2}-b^2\right)\right).
\label{PERT5}
\ee
In the limit of Lorentz invariance $a=b=1$ and we have $\aa_0=\aa$ and $\bb_0=\bb$.
The field renormalisation factor is
\be
Z^{1/2}=1+\fr{1}{3}\fr{g^2}{(4\pi)^2}\fr{1}{n-4}\left(\left(\fr{1}{b^2}+b^2\right)-5C_2(G)\right).
\ee

\subsection{\label{QSELFE0} Quark Self-Energy for Petrov Class O}

\begin{figure}
 \centering
 \includegraphics[width=0.3\linewidth]{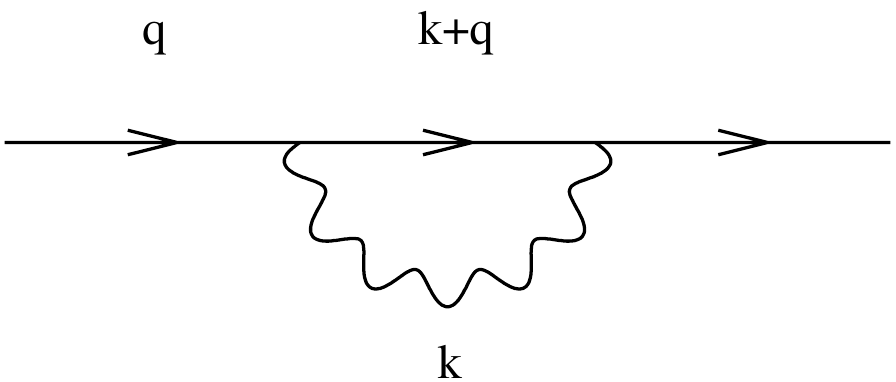}
 \caption{Quark self energy}
 \label{FIG5}
\end{figure}

The results for the calculations in the previous section, because each term involves only 
a single metric, can be read off from standard results (see reference \cite{PESK}). However  
the one loop self-energy correction to the quark propagator is obtained
from the Feynman rules applied to the diagram in Fig \ref{FIG5}. Both metrics are involved.
A possible conflict with causality might be anticipated. This possibility has been discussed 
previously and the conclusion is that in the present case there is no difficulty, as is
confirmed directly in the calculations \cite{ITD2,ITD3,SCHR,KLINK1,KLINK2}. We have then

\be
i\SS(p)=-g^2C_2(N)\int\fr{d^nk}{(2\pi)^n}\ggb^\mu\fr{1}{\ggb^\aa(p+k)_\aa-m}\ggb^\nu\fr{g_{\mu\nu}}{k^2}.
\ee
We require
\be
i\SS(0)=-g^2C_2(N)\int\fr{d^nk}{(2\pi)^n}\ggb^\mu\fr{1}{\ggb^\aa k_\aa-m}\ggb^\nu\fr{g_{\mu\nu}}{k^2},
\ee
and
\be
i\fr{\d}{\d p_\ll}\SS(p)|_{p=0}=i\SS^\ll(0)=g^2C_2(N)\int\fr{d^nk}{(2\pi)^n}\ggb^\mu\fr{1}{\ggb^\aa k_\aa-m}
                                     \ggb^\ll\fr{1}{\ggb^\bb k_\bb-m}\ggb^\nu\fr{g_{\mu\nu}}{k^2}.
\ee
The evaluation of these terms is along the same lines as the corresponding calculation for QED in \cite{ITD3}.
We find for the UV poles at $n=4$,
\be
i\SS(0)=im\fr{4g^2}{(4\pi)^2}C_2(N)\fr{1}{n-4}\fr{a+3b}{\sqrt{b}(\sqrt{a}+\sqrt{b})},
\ee
with the result
\be
\ss=\fr{4}{(4\pi)^2}C_2(N)\fr{1+3b^4}{b^2(1+b^2)}.
\label{RENORM29}
\ee
We have also
\be
i\SS^0(0)=i\fr{4g^2}{(4\pi)^2}C_2(N)\fr{1}{n-4}\fr{a-3b}{(\sqrt{a}+\sqrt{b})^2}\ggb^0,
\ee
\be
i\SS^j(0)=-i\fr{4}{3}\fr{g^2}{(4\pi)^2}C_2(N)\fr{1}{n-4}
                  \fr{(a+b)(2\sqrt{a}+\sqrt{b})}{\sqrt{b}(\sqrt{a}+\sqrt{b})^2}\ggb^j.
\ee
Using $a=b^{-3}$ we find for the mass renormalisation
\be
b^{(1)}=-\fr{4}{(4\pi)^2}C_2(N)\fr{1}{n-4}\fr{1+3b^4}{b^2(1+b^2)}.
\ee
We find also
\be
H=-\fr{4}{(4\pi)^2}C_2(N)\fr{2(1-b^2+2b^4)}{b^2(1+b^2)},
\label{RENORM30}
\ee
and
\be
h^\ll_{~~\rho}=\fr{4}{(4\pi)^2}C_2(N)\fr{(1-b^2)(1+3b^2+4b^4)}{2b^2(1+b^2)^2}\Tc^\ll_{~~\rho}.
\ee
Here $\Tc^\ll_{~~\rho}$ is a traceless diagonal matrix with entries $(1,-1/3,-1/3,-1/3)$.
It then follows that
\be
\bbb_0=\bbb\left(1+\fr{1}{3}\fr{4g^2}{(4\pi)^2}C_2(N)\fr{1}{n-4}\fr{(1-b^2)(1+3b^2+4b^4)}{b^2(1+b^2)^2}\right).
\ee
Combining this with eq(\ref{PERT5}) we obtain
\be
b_0=b\left(1+\fr{1}{3}\fr{g^2}{(4\pi)^2}\fr{1}{n-4}\fr{(1-b^2)}{b^2}\left[(1+b^2)+4C_2(N)\fr{(1+3b^2+4b^4)}{(1+b^2)^2}\right]\right).
\ee

\subsection{\label{COUPREN0} Coupling Constant Renormalisation for Petrov Class O}

\begin{figure}[t]
 \centering
 \includegraphics[width=0.4\linewidth]{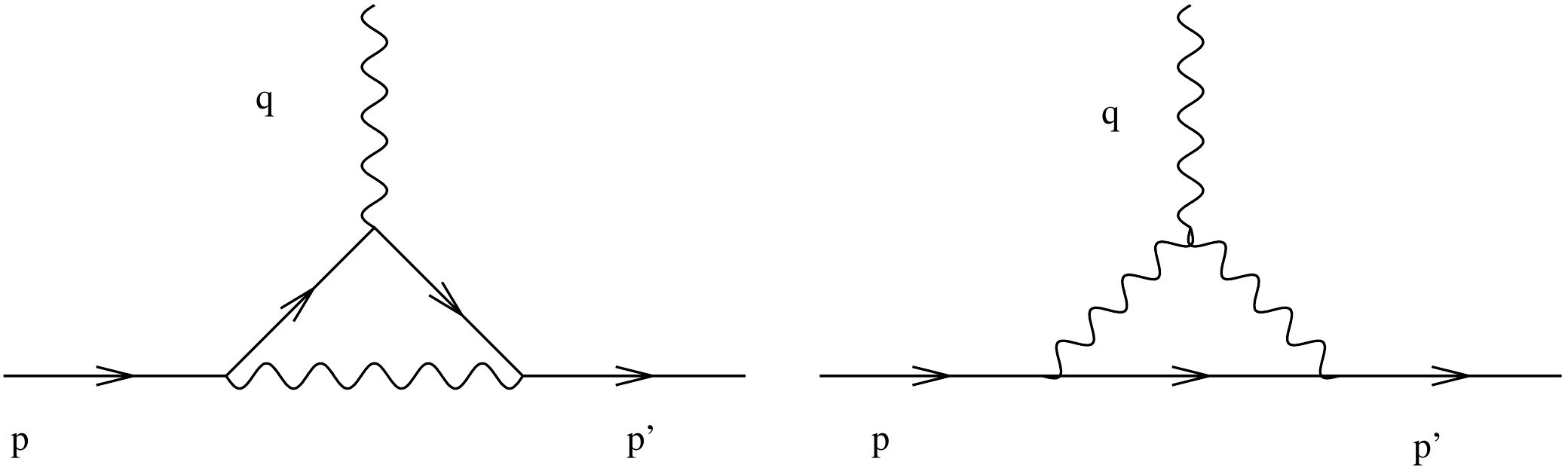}
 \caption{Quark gluon coupling diagram}
 \label{FIG6}
\end{figure}

The one loop diagrams in Fig \ref{FIG6} yield the coupling constant renormalisation. For the computation 
of the UV pole divergence it is sufficient to calculate the vertex with $p=p'=0$. From the first
diagram we obtain
\be
\Vc^\ll_a(0,0)=g^3t_bt_at_bg_{\mu\nu}I^{\mu\ll\nu},
\ee
where 
\be
I^{\mu\ll\nu}=\int \fr{d^nk}{(2\pi)^n}\ggb^\mu\fr{1}{\ggb^\aa k_\aa-m}
                          \ggb^\ll\fr{1}{\ggb^\bb k_\bb-m}\ggb^\nu\fr{1}{k^2}.
\ee
On omitting terms that do not contribute to the UV divergence we have
\be
I^{\mu\ll\nu}=\ggb^\mu\ggb^\aa\ggb^\ll\ggb^\bb\ggb^\nu T_{\aa\bb},
\ee
then
\be
T_{\aa\bb}=\int\fr{d^nk}{(2\pi)^n}\fr{k_\aa k_\bb}{(\gb^{\aa'\bb'}k_{\aa'}k_{\bb'}-m^2)^2k^2}
\label{CPRN1}
\ee
The result using $t_bt_at_b=(C_2(N)-C_2(G)/2)t_a$ and $ab^3=1$ is
\be
\Vc^0_a(0,0)=4\fr{ig^3}{(4\pi)^2}\fr{1}{n-4}(C_2(N)-C_2(G)/2)\fr{1-3b^4}{(1+b^2)^2}t_a\ggb^0.
\ee
We have also
\be
\Vc^j_a(0,0)=-\fr{4}{3}\fr{ig^3}{(4\pi)^2}\fr{1}{n-4}(C_2(N)_-C_2(G)/2)
                                                            \fr{(1+b^4)(2+b^2)}{b^2(1+b^2)^2}t_a\ggb^j.
\ee
In the case of the second diagram we note that in the limit of zero external momenta the internal 
three-gluon vertex reduces to

\be
V^{\ll\mu\nu}_{abc}=-gf_{abc}k_\rho(U^{\rho\nu\ll\mu}-U^{\rho\mu\nu\ll})
                       =-gf_{abc}k_\rho(2g^{\rho\ll}g^{\nu\mu}-g^{\rho\mu}g^{\nu\ll}-g^{\rho\nu}g^{\mu\ll}).
\ee
The contribution to the vertex becomes
\be
\Vc^\mu_a(0,0)=-\fr{1}{2}g^3t_a\ggb^{\nu'}\ggb^\aa\ggb^{\mu'}g_{\mu\mu'}g_{\nu\nu'}
                     (2g^{\rho\ll}g^{\mu\nu}-g^{\rho\mu}-g^{\rho\nu}g^{\mu\ll})\Tt_{\aa\rho},
\ee
where
\be
\Tt_{\aa\rho}=\int\fr{d^nk}{(2\pi)^4}\fr{k_\aa k_\rho}{(\gb^{\aa'\bb'}k_{\aa'}k_{\bb'}-m^2)(k^2)^2}.
\ee
We have again omitted terms that do not contribute to the UV pole at $n=4$. The tensor $\Tt_{\aa\rho}$
is closely related to $T_{\aa\rho}$ in eq(\ref{CPRN1}). Finally we have the contributions
\be
\Vc^0_a(0,0)=-4\fr{ig^3}{(4\pi)^2}C_2(G)\fr{1}{n-4}\fr{b^2(1+2b^2)}{(1+b^2)^2}t_a\ggb^0,
\ee
and
\be
\Vc^j_a(0,0)=-\fr{4}{3}\fr{ig^3}{(4\pi)^2}C_2(G)\fr{1}{n-4}\fr{1+2b^2+4b^4+2b^6 }{b^2(1+b^2)^2}t_a\ggb^j.
\ee
Combining the two sets of results we obtain
\be
\Vc^0_a(0,0)=4\fr{ig^3}{(4\pi)^2}\fr{1}{n-4}t_a\ggb^0\left(C_2(N)\fr{1-3b^4}{(1+b^2)^2}-\fr{1}{2}C_2(G)\right).
\ee
and
\be
\Vc^j_a(0,0)=-\fr{4}{3}\fr{ig^3}{(4\pi)^2}\fr{1}{n-4}t_a\ggb^j\left(C_2(N)\fr{(1+b^4)(2+b^2)}{b^2(1+b^2)^2}+\fr{3}{2}C_2(G)\right).
\ee
This leads to
\be
\Vc^\ll_a(0,0)=4\fr{ig^3}{(4\pi)^2}\fr{1}{n-4}t_aR^\ll_{~~\rho}\ggb^\rho,
\ee
where
\be
R^\ll_{~~\rho}=-\fr{1}{2}\left(C_2(N)\fr{1-b^2+2b^4}{b^2(1+b^2)}+C_2(G)\right)\dd^\ll_\rho
                                 +\fr{1}{2}C_2(N)\fr{(1-b^2)(1+3b^2+4b^4)}{b^2(1+b^2)^2}\Tc^\ll_{~~\rho}.
\ee
As expected from eq(\ref{RENORM25}) the second contribution to $R^\ll_{~~\rho}$ yields the correct term $\propto h^\ll_{~~\rho}$.
The bare coupling and the one loop correction yields
\be
\Vc^\ll_{0a}(0,0)+\Vc^\ll_a(0,0)=
                  i\mu^{(4-n)/2}g\left(1+g^{(1)}g^2-\fr{4g^2}{(4\pi)^2}\fr{1}{n-4}\fr{1}{2}\left(C_2(N)
                                       \fr{1-b^2+2b^4}{b^2(1+b^2)}+C_2(G)\right)\right)t_a\ggb^\ll
\label{CPRN2}
\ee
Thus from eq(\ref{RENORM28}) we find that
\be
K=-\fr{2}{(4\pi)^2}\left(C_2(N)\fr{1-b^2+2b^4}{b^2(1+b^2)}+C_2(G)\right).
\ee
The remaining UV divergences, as shown in eq(\ref{RENORM28}), are removed by appropriate field renormalisation factors and then finally 
by the pole in the coupling constant expansion. We have then from eq(\ref{CPRN2}) the result for the $\bb$-function
for the renormalised coupling
\be
\bb(g)=-(2-n/2)-\fr{g^3}{(4\pi)^2}\left(\fr{11}{3}C_2(G)-\fr{1}{3}\left(\fr{1}{b^2}+b^2\right)\right).
\ee 
Obviously the first term vanishes in four dimensions. The second term reduces to the standard answer
when $b=1$ and there is no Lorentz symmetry breaking. 

\subsection{\label{RENGRP0} Renormalisation Group for Petrov class O}

In four dimensions then, the important renormalisation group equations are for $g$ and $b$. They take the form
\be
\mu\fr{\d g}{\d\mu}=-\fr{g^3}{(4\pi)^2}\left(\fr{11}{3}C_2(G)-\fr{1}{3}\left(\fr{1}{b^2}+b^2\right)\right),
\label{RG4D_1}
\ee
and
\be
\mu\fr{\d b}{\d\mu}=-\fr{1}{3}\fr{g^2}{(4\pi)^2}\fr{1-b^2}{b^2}\left(1+b^2+4C_2(N)\fr{1+3b^2+4b^4}{(1+b^2)^2}\right).
\label{RG4D_2}
\ee
In general the two variables influence one another as they evolve along the RG trajectory. However
some points in $(g,b)$-space are particularly significant. The Lorentz invariant situation $b=1$ is stable 
and is maintained under the RG. The coupling constant $g$ then runs to zero, its fixed point, in the standard way as $\mu$ 
rises to infinity. The rate at which $g$ drops is the result of a competition between the contributions 
of the gauge field and the quark field to the vacuum polarisation. When we explore values of $b\ne 1$ we see 
that the effect of the quark field is enhanced with the result that $\bb(g)$ vanishes  when $b$ satisfies
\be
11C_2(G)-\fr{1}{b^2}-b^2=0.
\label{MINgi_1}
\ee
That is 
\be
b=b_{\pm}=\left(\fr{1}{2}(R\pm\sqrt{R^2-4})\right)^{1/2},
\label{MINg_2}
\ee
where $R=11C_2(G)$. On these two lines $\d b/\d\mu$ remains non-vanishing. The RG trajectories 
cross the lines and at the crossing point the coupling constant attains a minimum value. 
It increases again as the scaling energy $\mu$ continues to increase. 

If we modify the model so that it contains $n_f$ quarks, all sharing the same metric, 
then eq(\ref{RG4D_1}) becomes
\be
\mu\fr{\d g}{\d\mu}=-\fr{g^3}{(4\pi)^2}\left(\fr{11}{3}C_2(G)-\fr{n_f}{3}\left(\fr{1}{b^2}+b^2\right)\right),
\label{RG4D_3}
\ee
Eq(\ref{RG4D_2}) remains unchanged. The minimum of $g$ occurs when
\be
11C_2(G)-n_f(\fr{1}{b^2}+b^2)=0,
\label{MINg_3}
\ee
that is when  $b=b_{\pm}$ where now $R=11C_2(G)/n_f$. 

The RG equation for the quark mass is obtained from eq({\ref{RENGRP5}) and eq(\ref{RENORM29}).
It is
\be
\mu\fr{\d m}{\d\mu}=-2m\fr{g^2}{(4\pi)^2}C_2(N)\fr{1+b^2+4b^4}{b^2(1+b^2)}.
\label{RG4D4}
\ee

\section{\label{DISCUSSION} Discussion}

\begin{figure}[t]
 \centering
 \includegraphics[width=0.6\linewidth]{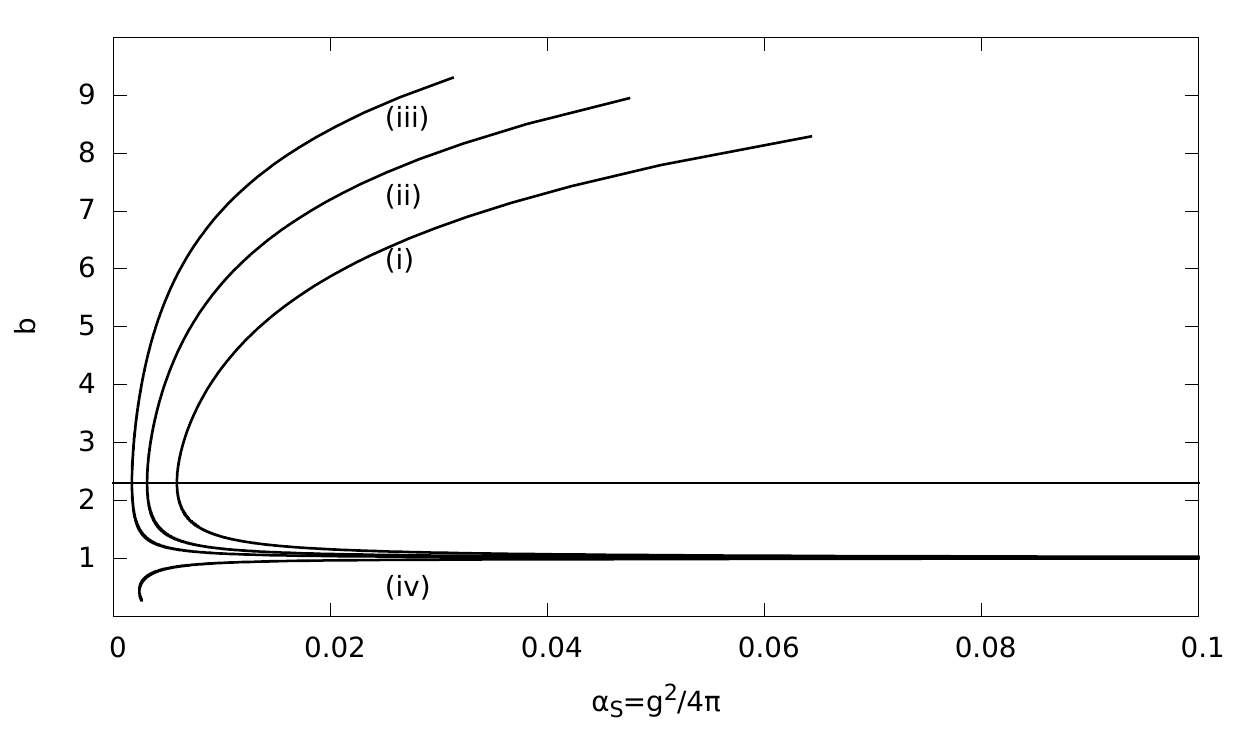}
 \caption{ The RG-trajectories for $SU(3)$ with $n_f=6$ starting at $\aa_S=0.1$.
The initial values of $b$ are (i) 1.025, (ii) 1.0125, (iii) 1.00625, (iv) 0.99375.
The horizontal line is at $b=b_+=2.3047$}
 \label{FIG7}
\end{figure}

The behaviour of the RG trajectory described above shows that a frustration of asymptotic freedom
can arise in the presence of LSV, at least in this model. The question arises as to whether
it might be observable experimentally. There are several issues to be considered, for example
the relationship of the lab frame to the gluon frame  used in the above discussion. We set this
matter provisionally aside, though it must ultimately be resolved, and assume that the lab frame
is travelling slowly relative to the gluon frame.
More significant is the size  of the energy range implicit in the model. We introduce
an initial energy $M_I$ and associated LSV parameter $b_I$, $QCD$ coupling $\aa_I=\aa_S(M_I)$
and the energy $M_{\hbox{min}}$ at which the coupling reaches its minimum value.

In a gesture towards "reality" we consider the case $SU(3)$ gauge theory, where $C_2(G)=3$ and 
$C_2(3)=4/3$ \cite {PESK} 
with $n_f=6$ quarks. In this case $b_+=2.3947$ and $b_-=0.4339$. The light cone of the quarks in the gluon 
frame at minimum coupling is $c_q=b_+^2=5.311$ for LSV with $b>1$ and $c_q=b_-^2=0.1882$ for LSV with $b<1$.
These values for $c_q$ represent rather severe LSV.

In order that  our asymptotic calculation be relevant we must assume $M_I$ is sufficiently large and in particular
is greater than the top quark mass, that is $173GeV$. In this energy regime there is no easy way
to relate our calculation to low energy determinations of $\aa_S(\mu)$. The story of the running coupling
and its relationship to low energy phenomena and $\LL_{QCD}$ is complicated. It is comprehensively
reviewed in \cite{DBT}. An important point is the subtraction scheme used to obtain finite results for
physical quantities. We are using the minimal subtraction ($MS$) scheme \cite{tHV}. More widely used 
is the $\overline{MS}$ scheme introduced in \cite{BBDM} to improve convergence. In that scheme the strong
coupling has an evaluation $\aa_S(M_Z\simeq 90 GeV)\simeq 0.12$. On the grounds that our calculation is 
exploratory we feel justified in neglecting the difference in subtraction schemes and propose $\aa_S(M_I)=0.1$ 
when $M_I=10^2-10^3 GeV$. The qualitative nature of the results is not altered by (relatively) small changes in 
our initial conditions.

With these initial conditions the results for examples of the renormalisation trajectory, obtained by 
numerical integration (2nd order Runge-Kutta) of eq(\ref{RG4D_3}) and eq(\ref{RG4D_2}) are shown in Fig \ref{FIG7}.
Obviously the closer the initial value $b_I$ is to unity the closer the renormalisation group                
trajectory stays near the Lorentz symmetry line and the later it breaks away, heading for its minimum value.
These results are illustrated in Fig \ref{FIG8} which shows the connection between 
$\log_{10}(M_{\hbox{min}}/M_I)$ and $b_I$. The smooth curve is obtained by fitting the rightmost point on
the plot. Even for this implausibly high value $b_I=1.1$ at our initial energy scale $M_I=10^2GeV$ we still 
find $M_{\hbox{min}}\simeq 10^{25}M_I$.
Tuning $b_I$ down to potentially more realistic values results in yet greater disparities in the orders of magnitude
of $M_{\hbox{min}}$ and experimentally attainable values for $M_I$. The conclusion must therefore be that for $QCD$ 
with the known set of quarks there is little hope of observing any of the frustration of asymptotic 
freedom in accelarator experiments. However the complex asymptotic behaviour that we encounter in this 
model may have relevance to very high energy processes at very early times in the initiating big bang of the universe.
In view of the fact that the energy range associated with frustration of aymptotic 
freedom appears to lie well above the Planck mass ($M_P \simeq 10^{19} GeV$) where gravitational effects must
become important, one might question its physical relevance. However it may also be possible and would certainly 
be interesting to relate the behaviour of $\aa_S(M)$ when $M\simeq M_P$ to models of quantum gravity
constructed with appropriate running LSV parameters \cite{EICH1,EICH2}.

These considerations do not preclude the possibility of discovering LSV effects 
in an energy range for which $b$ remains close to unity. For example if we set $b=1+x$ and assume 
$x$ is small then to lowest order in $x$ eq(\ref{RG4D_3}) and eq(\ref{RG4D_2}) become 
\be
\mu\fr{\d g}{\d\mu}=-A\fr{g^3}{(4\pi)^2},
\ee
\be
\mu\fr{\d x}{\d\mu}=B\fr{g^2}{(4\pi)^2}x,
\ee
where $A=(11C_2(G)-2n_f)/3$ and $B=4(1+4C_2(N))/3$. In this approximation 
\be
\aa_S(E)=\aa_I\left(1+7\fr{\aa_I}{4\pi}\log\fr{E}{M_I}\right)^{-1}.
\label{DISC}
\ee
the RG trajectories have the form
\be
b=1+x_I (\aa_S/\aa_I)^{-\kk},
\label{WFL}
\ee
where $x_I$ is the initial value of $x$ and $\kk=-B/2A$. Eq(\ref{WFL}) exhibits the instability at the fixed point
$(\aa_S,b)=(0,1)$. Depending on the sign selected for $x_I$, $b$ will either rise or fall from unity 
as $\aa_S$ approaches zero. With our choice of parameters we have $A=7$ and $B=8.4444$ 
with the result $\kk=0.6031$. The dependence of $x$ on $\aa_S$ is therefore relatively  weak. 
When $\aa_S$ decreases by an order of magnitude $x$ only increases by a factor of roughly 4. 
A similar approximation for the renormalised mass $m$ yields
\be
\fr{m}{m_I}=\left(\fr{\aa_S}{\aa_I}\right)^{\tau},
\ee 
where $\tau=0.762$. In the asymptotic energy range then, the renormalised mass reduces, also
relatively slowly, with a power of the renormalised coupling. If effective methods were developed for computing
the structure and scattering of high energy particles in the model (see references \cite{KOST5,LEHN2} for 
related discussions in QED and the Standard Model Extension) then possibly it could provide guidance for 
accelerator experiments and cosmic ray 
detectors investigating LSV phenomena in a high energy regime of $PeV$ and beyond. For example, 
if we take (intuitively) the quark metric, diag$(a,-b-b-b)$, as determining the dispersion relation 
for quark based states, it would become for a particle with mass $m$, energy $E$ and momentum $P$ 
\be
aE^2-bP^2=m^2c_q^4.
\ee
Combining the above results we find for the velocity of quark based particles
\be
v=\fr{d E}{d p}=\fr{p}{E}\left(1+Dx_I-F\fr{m_I^2}{E^2}\right),
\label{DISC2}
\ee
where
\be
D=4\left(\fr{\aa_S}{\aa_I}\right)^{-\kk}+\fr{14\kk\aa_I}{4\pi}\left(\fr{\aa_S}{\aa_I}\right)^{1-\kk}
\ee
and
\be
F=\fr{7\aa_I}{4\pi}\left(\fr{\aa_S}{\aa_I}\right)^{2\tau+1}\left(\tau+7(\tau-\kk)x_I\left(\fr{\aa_S}{\aa_I}\right)^{-\kk}\right)
\ee
The point here being that the coefficients in the dispersion relation depend only on $\aa_S/\aa_I$
and therefore vary only logarithmically with the energy $E$. The outcome for the velocity of quark based particles
shown in eq(\ref{DISC2}). Omitting all terms $O(\aa_I)$ that decrease logarithmically, we are left with the
simple result
\be
v\simeq\fr{p}{E}\left(1+4x_I\left(\fr{\aa_S}{\aa_I}\right)^{-\kk}\right)
\label{DISC3}
\ee
This is qualitatively different from LSV originating in higher derivative contributions to the 
QCD Lagrangian \cite{KOST7,KOST6} or spacetime foam models \cite{ELLIS1,ELLIS2}. These are parametrised 
by large mass scales and suggest powerlaw increases in energy. In our case eq(\ref{DISC3}) suggests 
a slow logarithmic increase that we might expect to be more difficult to detect. However were LSV to 
have been detected the suggested energy dependence would distinguish this QCD model from such higher derivative models.

\begin{figure}[t]
 \centering
 \includegraphics[width=0.6\linewidth]{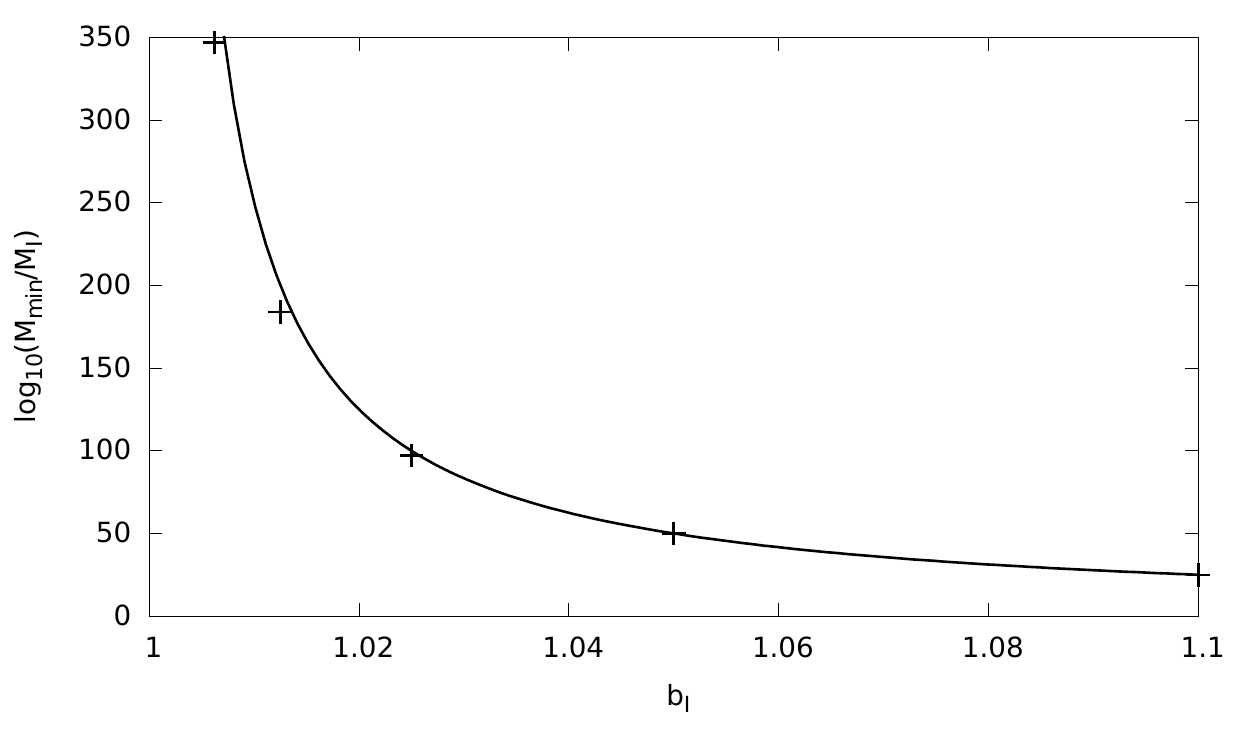}
 \caption{ Here $M_I$ is the initial energy scale and $b_I$ is the corresponding
LSV parameter, $M_{\hbox{min}}$ is the energy scale at which the running coupling attains a minimum.
The results of the R-K integration are represented by crosses and compare well with the continuous curve 
$\log_{10}(M_{\hbox{min}}/M_I)=2.5/(b_I-1)$.}
 \label{FIG8}
\end{figure}

\section{\label{CONC} Conclusions}

We have studied an $SU(N)$ QCD model with quarks in the fundamental representation 
and formulated the perturbation series to one loop with no restriction on the 
magnitude of the Lorentz symmetry breaking. In the particular case we studied the
LSV was due entirely to a mismatch between the lightcones of the quarks and gluons.
This is a consistent possibility if the lightcones are generated by two metrics 
that are both invariant under the same subgroup of the Lorentz group that leaves a 4-vector,
time-like in both metrics, invariant, a rotation group in fact. Similar results can be obtained 
with space-like and light-like vectors. 

The renormalisation group equation for the coupling constant $\aa_S$ and the LSV parameter $b$
was obtained  with the result, exhibited in Fig\ref{FIG7}, that initially $\aa_S$ 
decreases with energy just as in the standard Lorentz symmetric case. However
$b$ departs from unity increasingly with energy and this enhances the contribution of the quark vacuum
polarisation to the $\bb$-function for $\aa_S$. The outcome is that at sufficiently high energy
$\aa_S$ ceases to decrease, reaches a minimum and then increases again with energy. This
constitutes the frustration of asymptotic freedom in QCD with LSV of the kind we have investigated. 
We suggest plausible values for the energy range $E>M_I$  we are investigating and the associated
initial value $\aa_I$ for the strong coupling. The outcome is that the frustration part of the
RG trajectory for $(\aa_S,b)$ is at energies many orders of magnitude greater than is accessible to
accelerator experiments. It is well above the Planck mass. However it is possible that part of the 
RG trajectory lying near the Lorentz symmetry line $b=1$ might be attainable in accelerator
or cosmic ray observations. The effect on the dispersion relation of particles is through
powers of $\aa_S$ and hence is logarithmic in character and represents a kind of intrinsic
LSV rather than one parametrised by higher derivative contributions to the Lagrangian.

There are many variations of the model that might be investigated such as increasing the 
number of quarks, varying the quark metrics in ways that induce more complex LSV
associated with higher Petrov classes. Of course one should also consider how these
results relate to the full structure of the Standard Model and its extensions.
Finally it is worth noting that in the context of relatively weak LSV it may be possible 
to pursue a nonperturbative investigation of our model using the techniques of lattice QCD.

\section*{Acknowledgements}

This work has been partially supported by STFC consolidated grant ST/P000681/1. 
I am grateful to R. R. Horgan for discussions concerning the potential relevance 
of lattice field theory calculations to evaluating LSV in gauge theories.

\bibliography{mm2}
\bibliographystyle{unsrt}

\end{document}